\documentstyle[aps]{revtex}
    \font\tenrm=cmr10
\title{Distributional Sources in General Relativity: two point-like
examples revisited}
\author{\it
    {\tenrm N. R. PANTOJA}\\
    Centro de Astrof\'{\i}sica Te\'{o}rica.\\
    Departamento de F\'{\i}sica, Facultad de Ciencias.
    Universidad de los Andes.\\
    M\'{e}rida, 5101, Venezuela.\\
    \bigskip
    {\tenrm H. RAGO}\\ 
    Laboratorio de F\'{\i}sica Te\'{o}rica and
    Centro de Astrof\'{\i}sica Te\'{o}rica.\\
    Departamento de F\'{\i}sica, Facultad de Ciencias.
    Universidad de los Andes.\\
    M\'{e}rida, 5101, Venezuela.}
\begin{document}
\maketitle
    \newcommand{\ttensor}[2]{\mbox{$\left(\parbox{3ex}{\centerline{#1}\par \centerline{#2}}\right)$}-tensor}

\begin{abstract}
A regularization procedure, that allows one to relate
singularities of curvature to those of the Einstein tensor without
some of the shortcomings of previous approaches, is proposed. This
regularization is obtained by requiring that (i) the density $|det
{\bf g}|^{\frac{1}{2}}G^a_{\,b}$, associated to the Einstein
tensor $G^a_{\,b}$ of the regularized metric, rather than the
Einstein tensor itself, be a distribution and (ii) the regularized
metric be a continuous metric with a discontinuous extrinsic
curvature across a non-null hypersurface of codimension one. In
this paper, the curvature and Einstein tensors of the geometries
associated to point sources in the $2+1$-dimensional gravity and
the Schwarzschild spacetime are considered. In both examples the
regularized metrics are continuous regular metrics, as
defined by Geroch and Traschen, with well defined distributional
curvature tensors at all the intermediate steps of the
calculation. The limit in which the support of these curvature
tensors tends to the singular region of the original spacetime is
studied and the results are contrasted with the ones obtained in
previous works.
\end{abstract}

\pacs{PACS numbers: 0250N, 9760L}

\section{Introduction}
Within the framework of general relativity, a spacetime singularity
corresponds to a singularity of the metric tensor which can not be
removed by a coordinate transformation, even one which itself becomes
singular where the metric does. On the other hand, it is expected
that this singularity reveals itself through a lack of smoothness of
the curvature tensor. In order to incorporate singular curvatures
into general relativity, distributional curvatures have been
considered in several papers\cite{parker,taub,raju,geroch}. However,
as is well known, computing the curvature tensor from a metric
requires nonlinear operations which are not defined within the
framework of distribution theory. This imposes strong constraints on
the class of metrics whose curvature tensors make sense as
distributions.

There is a class of metrics, the regular metrics\cite{geroch}, for
which the curvature tensor has a well defined distributional meaning.
The singular parts of the curvature tensor of these regular metrics
for a $d$-dimensional spacetime are supported on submanifolds of
codimension of at most one, i.e., of dimension $\geq d-1$, and for
these metrics it makes sense to write Einstein's equations with
energy-momentum tensor distributions. Metrics for surface
layers\cite{israel}, turn out to be included into the class of
regular metrics\cite{geroch,khorrami91,khorrami94,mansouri96}.
However, there are some very simple metrics, from which a physically
interesting spacetime follows, which are not regular. One example is
the $3+1$-dimensional Minkowski metric with an angular
deficit\cite{geroch,garfinkle}, another is the Schwarzschild
metric\cite{geroch}. Regular metrics are a subclass of a larger class
of metrics, semi-regular metrics\cite{garfinkle}. The curvature of a
semi-regular metric is well defined as a distribution and it has been
proved that the $3+1$-dimensional Minkowski metric with an angular
deficit and a certain kind of traveling wave metric are semi-regular
metrics\cite{garfinkle}. Alternatively, by considering the
Colombeau's theory of generalized functions\cite{colombeau},
distributional curvatures can be defined. This approach has been used
to obtain the distributional curvature associated to a conical
singularity \cite{clarke,wilson,vickers} but we will not consider it
here.

By invoking regularization procedures, a distributional meaning may
be given to the curvature tensor of a non-regular metric. However,
the reference differentiable structure that the regularization
implicitly uses, makes uncertain the independence of the result on
the regularization procedure chosen\cite{louko}. It is well known
that problems appear in defining the curvature tensor for the
$3+1$-dimensional Minkowski spacetime with an angular deficit, due to
the fact that the distributional limit depends on the choice of
regularization\cite{geroch}. Regularizations of the metric have been
also used to calculate distributional curvatures for the
Schwarzschild\cite{balasin93,kawai} and Kerr-Newman
spacetimes\cite{balasin94}, with results which are regularization
scheme dependent\cite{kawai}.

In this work we propose a very restrictive kind of regularization
inspired by the approach used in early works to the study of the
classical gravitational self-energy and the minimum extension
associated to point-like sources in general
relativity\cite{arnowitt60,arnowitt65}. We show that this approach
may be used to regularize non-regular metrics in such a way that the
regularized metrics are continuous regular metrics with well defined
distributional curvature tensors, in the sense of
reference\cite{geroch}, at every intermediate step in the
calculation. Furthermore, since continuous regular metrics can be
suitably approximated by smooth metrics\cite{geroch}, this approach
provides a physically sensible idealization for the smooth metrics of
general relativity.

In section \ref{sec2}, after an overview on the subject of
distributions on metric manifolds, we review the definitions of
regular and semi-regular metrics. In the following section, the
distributional curvature and Einstein tensors of the
$(2+1)$-dimensional spacetime around a massive point
source\cite{deser84} are considered in some detail. First,
following Ref.\cite{garfinkle}, we show that the metric of this
spacetime is a semi-regular metric with a well defined
distributional curvature but with an Einstein tensor which is zero
everywhere. Next, we regularize the metric by requiring that (i)
the tensor density $|det {\bf g}|^{\frac{1}{2}}G^{a}_{\,b}$
associated to the mixed-index Einstein tensor $G^{a}_{\,b}$,
rather than the Einstein tensor itself, be a well defined
distribution and (ii) the regularized metric be a continuous
metric with a discontinuous extrinsic curvature across a non-null
hypersurface of codimension one. The distributional limit to a
point source is shown to be in agreement with the identification
made in reference \cite{deser84}. Finally, we show that the
regularized metric is a regular metric and, following the approach
of Ref.\cite{geroch}, calculate the curvature and Einstein
tensors. The distributional limit in which the regularization is
removed reveals the origin of the disagreement between the results
obtained. In section \ref{sec4}, the Schwarzschild spacetime is
considered. We first prove that the Schwarzschild metric is not a
semi-regular metric and show some drawbacks of previous
regularization approaches. Next, following the regularization
procedure of section \ref{sec3}, we regularize the Schwarzschild
metric obtaining a regular metric. The results, in the limit in
which the regularization is removed, are shown to be in agreement
with those obtained in previous works. Finally, following the
approach of Ref.\cite{geroch}, we calculate the curvature and
Einstein tensors of the regularized metric. The results obtained
by taking the limit in which the regularization is removed are
then contrasted with the previous ones.

\section{Regular and semi-regular metrics}\label{sec2}
Let us briefly review the class of metrics which have been defined
as regular and semi-regular metrics. For this, we first recall
some fundamental results about distributions. Since we shall have
no need for the theory of distributions on arbitrary manifolds in
its greatest generality\cite{choquet,lichnerowicz}, we shall
simplify wherever possible.

Let $\Phi({\cal M})$ be the family of $C^{\infty}$ real-valued
scalar functions $\phi$ defined on an orientable $n$-dimensional
$C^{\infty}$ paracompact manifold ${\cal M}$ and vanishing outside
some compact region of ${\cal M}$. Further, suppose that a rule
can be introduced in $\Phi({\cal M})$ defining the convergence to
zero of a sequence of functions $\phi_n\, (n=1,2,\ldots)$
belonging to $\Phi({\cal M})$.

Let ${\bf g}$ be a metric on ${\cal M}$ with associated volume
element ${\bf \omega_{{\bf g}}}= |det {\bf g}|^{\frac{1}{2}}
{\bf \varepsilon}$, where ${\bf \varepsilon}= {\bf
d}x^1\wedge\ldots\wedge{\bf d}x^n$ is the associated coordinate
volume element. We can define the functional generated by the
$0$-form $f$ on $\Phi({\cal M})$ through the rule
\begin{equation}
\int_{\cal M} {\bf \omega}_{{\bf g}} f \phi= \int_{U_{\cal M}}
dx^1\ldots dx^n \, |det {\bf g}|^{\frac{1}{2}} f(x)\phi(x),
\label{primer}
\end{equation}
where $U_{\cal M}$ is the coordinate domain corresponding to
${\cal M}$. Obviously, this identification does not depend on the
choice of coordinate system covering the corresponding domain, but
depends on the choice of ${\bf g}$ through the volume element
${\bf \omega_{{\bf g}}}$ .

Let ${\bf \eta}$ be a $C^{\infty}$ metric tensor. If we endow the
manifold ${\cal M}$ with such a metric, every locally integrable
$0$-form $f$ defines a distribution through
(\ref{primer}), and write
\begin{equation}
f[\phi]=\int_{\cal M} {\bf \omega}_{{\bf \eta}} f \phi.
\end{equation}
However, it should be noted that Eq.(\ref{primer}) allows one to
consider the functional generated by the scalar density $|det {\bf
g}|^{\frac{1}{2}}f$, even when ${\bf g}$ is a $C^k$
metric tensor.

The extension to tensor distributions is straightforward. Let
${\bf U}$ be a smooth \ttensor{q}{p} with compact support on
${\cal M}$. A locally integrable \ttensor{p}{q} field ${\bf T}$ is
identified with a tensor distribution via
\begin{equation}
{\bf T}[{\bf U}]\equiv \int_{\cal M} ({\bf T} |{\bf U})\omega_{\eta},
\label{Tdist}
\end{equation}
where
\begin{equation}
({\bf T}|{\bf U})\equiv T^{i_1\ldots i_p}\!_{j_1\ldots
j_q}U^{j_1\ldots j_q}\!_{i_1\ldots i_p}.\label{sd}
\end{equation}
A \ttensor{p}{q} field ${\bf T}$ is called locally bounded
provided that (\ref{sd}) is bounded for all test \ttensor{q}{p}
fields ${\bf U}$.

The derivative in the smooth metric ${\bf \eta}$ of a \ttensor{p}{q}
distribution ${\bf T}$ is the \ttensor{p}{q+1} defined for every test
\ttensor{q+1}{p} ${\bf U}$ by
\begin{equation}
{\bf \nabla} {\bf T}[{\bf U}]= -{\bf T}[\eta \cdot {\bf \nabla} {\bf
U}],
\end{equation}
where
\begin{equation}
(\eta \cdot {\bf \nabla} {\bf U})^{j_1\ldots j_q}\!_{i_1\ldots i_p}=
\nabla_j U^{jj_1\ldots j_q}\!_{i_1\ldots i_p}.
\end{equation}

Finally, the weak derivative of a locally integrable \ttensor{p}{q}
tensor field ${\bf T}$ is a locally integrable \ttensor{p}{q+1} field
${\bf W}$, if one exist, such that
\begin{equation}
{\bf W}[{\bf U}]= {\bf \nabla}{\bf T}[{\bf U}],
\end{equation}
for every test \ttensor{q+1}{p} field ${\bf U}$. For an approach in
which tensor distributions and their derivatives are described
without assuming the presence of a metric see reference \cite{dray}.
Alternatively, by using de Rham currents\cite{rham} and replacing
test tensors by test $n$-forms, the introduction of a volume element
can be avoided. Here we shall follow the more conventional approach
of reference\cite{geroch}.

Now, let $\nabla$ be the derivative operator in any smooth metric
${\bf \eta}$. Then the Riemann tensor of a smooth metric ${\bf g}$
can be written as
\begin{equation}
R_{abc}^{\quad d}= {\tilde{R}}_{abc}^{\quad d} +
2\nabla_{[b}C^d_{a]c} + 2 C^d_{m[b}C^m_{a]c},\label{RR}
\end{equation}
where
\begin{equation}
C^c_{ab} \equiv \frac{1}{2}(g^{-1})^{cd}(\nabla_a g_{bd} +
\nabla_b g_{ad} - \nabla_d g_{ab}) \label{chris})
\end{equation}
and ${\tilde{R}}_{abc}^{\quad d}$ is the curvature tensor of ${\bf
\eta}$. For the Einstein tensor of ${\bf g}$ we have
\begin{equation}
G_{ab}= R_{ab} - \frac{1}{2}(g^{-1})^{cd}{\tilde{R}}_{cd}g_{ab} +
(g^{-1})^{cd}C^e_{m[c}C^m_{e]d}g_{ab} +
\nabla_{[c}\left(C^e_{e]d}(g^{-1})^{cd}g_{ab}\right) +
C^e_{d[c}\nabla_{e]}\left( (g^{-1})^{cd}g_{ab}\right),\label{ER}
\end{equation}
where $R_{ab}= R_{acb}^{\quad c}$ is the Ricci tensor.

Following Ref.\cite{geroch}, a symmetric tensor field ${\bf g}$ on a
manifold ${\cal M}$ will be called a regular metric provided: (i)
${\bf g}$ and ${\bf g}^{-1}$ exist everywhere and are locally
bounded, and (ii) the weak derivative of ${\bf g}$ in some smooth
metric ${\bf \eta}$ exists and is locally square-integrable, i.e.,
the outer product of the weak derivative with itself is locally
integrable. The curvature tensor (\ref{RR}) and the Einstein tensor
(\ref{ER}) of a regular metric make sense as distributions, therefore
it makes sense to write Einstein's equations with distributional
energy-momentum tensors. It turns out that these idealized matter
sources must be concentrated on submanifolds of codimension of at
most one\cite{geroch}.

A wider class of metrics whose curvature makes sense as a
distribution can be defined, the so-called semi-regular metrics
\cite{garfinkle}. A symmetric tensor field {\bf g} on a manifold
${\cal M}$ will be called a semi-regular metric provided: (i)
${\bf g}$ and ${\bf g}^{-1}$ exist almost everywhere and are
locally integrable, and (ii) the weak first derivative $\nabla
{\bf g}$ of ${\bf g}$ in some smooth metric ${\bf \eta}$ exist and
the tensors $C^c_{ab}$ and $C^d_{m[b}C^m_{a]c}$ are locally
integrable. Under these constraints, the right hand side of
(\ref{RR}) is well defined as a distribution. However, a
semi-regular metric may have no distributional Einstein tensor.
This is due to the fact that for a semi-regular metric,
contractions of the metric with the curvature tensor may have no
sense as distributions.

\section{Point sources in $(2+1)$-dimensional gravity}\label{sec3}
Consider a $(2+1)$-dimensional spacetime $({\cal R}^3,{\bf g})$,
where the metric ${\bf g}$ in a particular coordinate system is
given by
\begin{equation}
{\bf g}
    = -{\bf d}t\otimes{\bf d}t +\rho^{-8m}({\bf d}\rho\otimes{\bf d}\rho +
    \rho ^2{\bf d}\varphi\otimes{\bf d}\varphi),
    \label{cero}
\end{equation}
where $-\infty < t < \infty$, $0 < \rho$ and $-\pi \leq \varphi
\leq \pi$, with the surfaces $\varphi = -\pi, \pi$ identified. In
Ref.\cite{deser84}, the metric (\ref{cero}) has been identified
with the metric generated by a point source of mass $m$ at the
origin.

It is easy to see that (\ref{cero}) is not a regular metric (the
analogous treatment of the problem in $(3+1)$-dimensional gravity
has been considered elsewhere \cite{geroch,garfinkle}). Let ${\bf
\eta}$ be the ordinary Minkowski metric on ${\cal R}^3$ given by
\begin{equation}
{\bf \eta}
    = -{\bf d}t\otimes{\bf d}t +{\bf d}\rho\otimes{\bf d}\rho +
    \rho ^2{\bf d}\varphi\otimes{\bf d}\varphi,
    \label{minkowski}
\end{equation}
and we take for the differentiable structure that in which $t,
x=\rho \cos \varphi$ and $y=\rho \sin \varphi$ form a smooth
chart. It follows that
\begin{equation}
{\bf g}= {\bf \eta} - (1-\rho^{-8m})({\bf d}\rho\otimes{\bf d}\rho
+ \rho^2{\bf d}\varphi\otimes{\bf d}\varphi)
\end{equation}
and
\begin{equation}
{\bf g}^{-1}= {\bf \eta}^{-1} - (1-\rho^{8m}) (
\partial_{\rho}\otimes
\partial_ {\rho} +
\rho^{-2}\partial_{\varphi}\otimes
\partial_ {\varphi}).
\end{equation}
We have that ${\bf g}$ and ${\bf g}^{-1}$ exist almost everywhere.
Let ${\bf U}$ be a test \ttensor{2}{0} field on ${\cal R}^3$. We
have
\begin{equation}
{\bf g}[{\bf U}]= \int_{{\cal R}^3}({\bf \eta}|{\bf
U})\omega_{\eta} - \int_{{\cal R}^3}(1-\rho^{-8m})(U^{xx} +
U^{yy})\omega_{\eta}.
\end{equation}
Therefore, ${\bf g}$ is locally integrable for $0\leq 8m < 2$ but
not locally bounded. Next, let ${\bf S}$ be a test \ttensor{0}{2}
field on ${\cal R}^3$. It follows that
\begin{equation}
{\bf g}^{-1}[{\bf S}]= \int_{{\cal R}^3}({\bf S}|{\bf
\eta}^{-1})\omega_{\eta} - \int_{{\cal R}^3}(1-\rho^{8m})(S_{xx} +
S_{yy})\omega_{\eta}.
\end{equation}
Hence, ${\bf g}^{-1}$ is locally bounded and locally integrable for
all $m \geq 0$. Finally, let ${\bf U}$ be a test \ttensor{3}{0} field
on ${\cal R}^3$. We find that the weak derivative in ${\bf \eta}$ of
${\bf g}$ exist almost everywhere and is given by
\begin{equation}
{\bf \nabla} {\bf g}[{\bf U}]= -{\bf g}[\eta \cdot {\bf \nabla}
{\bf U}]= \lim_{\epsilon \rightarrow 0} \int_{\rho >\epsilon}
g_{ab}\nabla_c U^{cab} \omega_{\eta}= \int_{{\cal R}^3}
W_{cab}U^{cab}\omega_{\eta},
\end{equation}
where
\begin{equation}
W_{cab}= -8m \rho^{-(1+8m)}d\rho_c(d\rho_ad\rho_b +
    \rho ^2d\varphi_a d\varphi_b).\label{weak}
\end{equation}
It then follows that ${\bf \nabla}{\bf g}$ is not locally square
integrable for $m > 0$. Therefore, ${\bf g}$ is not a regular
metric in the differentiable structure chosen. Actually, there
exists no differentiable structure that will render the metric
(\ref{cero}) regular since the support of the curvature of
(\ref{cero}) is expected to be a submanifold of codimension
greater than one.

Now, from (\ref{cero}) and (\ref{chris}) it follows that
\begin{equation}
C^c_{ab}= \frac{4m}{\rho}\left( \partial_{\rho}^c  d\rho_ad\rho_b
+ \partial_{\varphi}^c(d\rho_a d\varphi_b + d\rho_b d\varphi_a)
-\rho^2
\partial_{\rho}^c  d\varphi_a  d\varphi_b
\right),\label{chriscon}
\end{equation}
which is locally integrable. On the other hand
\begin{equation}
C^d_{m[b}C^m_{a]c}=0,\label{CC}
\end{equation}
which is also locally integrable. Therefore, the metric
(\ref{cero}) is a semi-regular metric.

From (\ref{RR}), (\ref{chriscon}) and (\ref{CC}), the Ricci tensor
is given by
\begin{equation}
R_{ab}= \nabla_{[c}C^c_{a]b}.
\end{equation}
Hence
\begin{equation}
R_{ab}[S^{ab}]= -\int_{{\cal R}^3} \left( C^c_{ab}\nabla_cS^{ab} -
C^c_{cb}\nabla_aS^{ab}\right) \omega_{\eta}
\end{equation}
and we obtain
\begin{eqnarray}
R_{ab}[S^{ab}]&=& - \lim_{\epsilon \rightarrow 0} \int_{\rho
>\epsilon}\left( C^c_{ab}\nabla_cS^{ab} -
C^c_{cb}\nabla_aS^{ab}\right) \omega_{\eta}\nonumber\\ &=&
\lim_{\epsilon \rightarrow 0} \left(\int_{\rho
=\epsilon}(\nabla_c\rho C^c_{ab} - \nabla_a\rho
C^c_{cb})S^{ab}\sigma + \int_{\rho
>\epsilon}(\nabla_c C^c_{ab} -
\nabla_a C^c_{cb})S^{ab}\omega_{\eta}\right),
\end{eqnarray}
where $\sigma$ is the volume element induced on the surface $\rho=
constant$ by the metric (\ref{minkowski}). For $\rho \neq 0$ we
have
\begin{equation}
\nabla_c C^c_{ab}= \nabla_a C^c_{cb}= \frac{8m}{\rho^2}(d\rho_a
d\rho_b -
    \rho ^2 d\varphi_a d\varphi_b).
\end{equation}
Then
\begin{eqnarray}
R_{ab}[S^{ab}]&=& - \lim_{\epsilon \rightarrow 0} \int_{\rho
=\epsilon}(d\rho_c C^c_{ab} -
d\rho_aC^c_{cb})S^{ab}\sigma\nonumber
\\ &=&  \lim_{\epsilon \rightarrow 0} \int_{\rho
=\epsilon}\frac{4m}{\rho}(d\rho_a d\rho_b +
    \rho ^2 d\varphi_a d\varphi_b)S^{ab} \sigma \nonumber\\
&=&   \lim_{\epsilon \rightarrow 0} 4m \int_{\rho =\epsilon}dt
d\varphi \, (dx_a dx_b + dy_a dy_b) S^{ab}\nonumber\\ &=& 8\pi m
\int dt\,\left(S^{xx}(t,0,0) +
S^{yy}(t,0,0)\right).\label{Riccicalc}
\end{eqnarray}
Therefore
\begin{equation}
R_{ab}= 8\pi m \delta^{(2)}_{(0)}(dx_a dx_b +dy_a
dy_b).\label{Riccifinal}
\end{equation}

Next, from (\ref{ER}),(\ref{weak}) and (\ref{chriscon}), the
Einstein tensor of (\ref{cero}) is given by
\begin{equation}
G_{ab}= R_{ab} -8m\nabla_c\left( -\rho^{8m-1} \partial_{\rho}^c
dt_a dt_b +\rho^{-1}(d\rho_a d\rho_b + \rho^2 d\varphi_a
d\varphi_b)\right) - 64m^2\rho^{8m-2}dt_a dt_b.\label{sorp}
\end{equation}
Note that the right hand side of (\ref{sorp}) contains the
derivative of a locally integrable tensor plus a locally
integrable tensor. Hence, the Einstein tensor of (\ref{cero})
makes sense as a distribution. Then
\begin{eqnarray}
G_{ab}[S^{ab}]=&& R_{ab}[S^{ab}]\nonumber\\ &+& 8m\int_{{\cal
R}^3}\left( -\rho^{8m-1} \partial_{\rho}^c dt_a dt_b
+\rho^{-1}(d\rho_a d\rho_b + \rho^2 d\varphi_a
d\varphi_b)\right)\nabla_cS^{ab}\omega_{\eta}\nonumber\\
&-&64m^2\int_{{\cal R}^3}\rho^{8m-2}dt_a dt_b S^{ab}
\omega_{\eta}.\label{sorpdis}
\end{eqnarray}
An analogous calculation to that of (\ref{Riccicalc}) leads to
\begin{eqnarray}
\int_{{\cal R}^3}\left( -\rho^{8m-1} \partial_{\rho}^c dt_a dt_b
\right. &+& \left.\rho^{-1}(d\rho_a d\rho_b + \rho^2 d\varphi_a
d\varphi_b)\right)\nabla_cS^{ab}\omega_{\eta} = \nonumber\\ &-&
\pi \delta^{(2)}_{(0)}(dx_a dx_b + dy_a dy_b)[S^{ab}] +
8m\int_{{\cal R}^3}\rho^{8m-2}dt_a dt_b S^{ab}
\omega_{\eta}.\label{inter}
\end{eqnarray}
It follows from (\ref{Riccifinal}),(\ref{sorpdis}) and
(\ref{inter}) that
\begin{equation}
G_{ab}=0.\label{Gsemreg}
\end{equation}
These results suggest that the spacetime $({\cal R}^3,{\bf g})$
with ${\bf g}$ given by (\ref{cero}), although having a
distributional curvature with support on the origin, has a zero
everywhere distributional Einstein tensor. How are we to reconcile
these results with those of Ref.\cite{deser84}?

In the following, the approach of references
\cite{arnowitt60,arnowitt65} is considered. We show that this
approach may be used to regularize (\ref{cero}) in such a way that
the resulting regularized metric is a continuous regular metric
with well defined distributional curvature and Einstein tensors in
the sense of reference \cite{geroch}.

Consider a 3-dimensional spacetime $({\cal R}^3,{\bf g})$, where
the metric ${\bf g}$ is given by
\begin{equation}
{\bf g}
    = g_{\alpha\beta}{\bf d}x^\alpha\otimes {\bf d}x^\beta
    = -{\bf d}t\otimes{\bf d}t +H({\bf d}\rho\otimes{\bf d}\rho +
    \rho ^2{\bf d}\varphi\otimes{\bf d}\varphi),
    \label{uno}
\end{equation}
where $-\infty < t < \infty$, $0 \leq \rho$ and $-\pi \leq \varphi
\leq \pi$, with the surfaces $\varphi = -\pi, \pi$ identified. At
this stage, $H=H(\rho)$ is an unknown $C^{\infty}$ function,
making possible to perform conventional pointwise differential
geometry.

From (\ref{uno}), it follows that the curvature two-form, ${\cal
R}_{\rho}^\varphi$, is given by
\begin{equation}
{\cal R}_{\rho}^\varphi
    = \frac{1}{2}\frac{d}{d\rho}(\rho\frac{d}{d\rho}\log H)
    {\bf d}\rho\wedge {\bf d}\varphi
\end{equation}
and the non-zero components of the Riemann tensor are
\begin{equation}
R_{\quad\rho\varphi}^{\rho\varphi}
    = R_{\quad\varphi\rho}^{\varphi\rho}
    = -\frac{1}{2\rho H}\frac{d}{d\rho}(\rho\frac{d}{d\rho}\log H).
   \label{cinco}
\end{equation}
From (\ref{cinco}), the Ricci tensor and the scalar curvature can be
obtained. Finally, the Einstein tensor ${\bf G}$ is found to be given by
\begin{equation}
{\bf G}= G^t_t {\bf d}t\otimes\partial_t,\label{einsteina}
\end{equation}
where
\begin{equation}
G^t_t= \frac{1}{2\rho H}\frac{d}{d\rho}(\rho\frac{d}{d\rho}\log
H). \label{einstein}
\end{equation}
Here, it is worth recalling that in three spacetime dimensions the
identity
\begin{equation}
R^{\mu\nu}_{\alpha\beta} =
    \varepsilon^{\mu\nu\sigma}
    \varepsilon_{\alpha\beta\lambda}G^{\lambda}_{\sigma}
    \label{GR}
\end{equation}
holds, linking curvature and Einstein tensors.

Now, we take (\ref{einsteina},\ref{einstein}) as the definition of
the Einstein tensor of (\ref{uno}). Let ${\bf U}$ be a test
\ttensor{1}{1} field on ${\cal R}^3$.  Next, consider the
mixed-index Einstein tensor density $|det {\bf
g}|^{\frac{1}{2}}{\bf G}$, with $|det {\bf g}|^{\frac{1}{2}}$ the
density generated by the metric ${\bf g}$ in (\ref{uno}), as a
functional on the space of test tensor fields through
\begin{equation}
  \int_{{\cal R}^3} |det {\bf
  g}|^{\frac{1}{2}}G^{a}_{\,b}U^{b}_{\,a}{\bf \varepsilon}
  = \int_{{\cal R}^3} dt dx dy \ H G^{t}_{\,t}U^{t}_{\,t},
\end{equation}
where it is understood that $G^{a}_{\,b}$ and $U^{b}_{\,a}$ are
Cartesian components as functions of Cartesian coordinates
$(t,x,y)$ with $x=\rho\cos\phi$ and $y=\rho\sin\phi$. Then
\begin{equation}\label{disteins}
  \int_{{\cal R}^3} |det {\bf
  g}|^{\frac{1}{2}}G^{a}_{\,b}U^{b}_{\,a}{\bf \varepsilon}=
  \lim_{\epsilon \rightarrow 0}\int_{\rho > \epsilon}dt d\rho
d\varphi\, \frac{1}{2} \frac{d}{d\rho}\left(\rho \frac{d}{d\rho}\log
H \right) U^{t}_{\,t}(t,\rho \cos \varphi,\rho \sin \varphi) =
    \lim_{\epsilon \rightarrow 0} \pi\int_{\epsilon}^{\infty}
    d\rho \, \frac{d}{d\rho} \left(\rho \frac{d}{d\rho}\log
H \right) \overline{U^{t}_{\,t}}(\rho),
\end{equation}
where
\begin{equation}
  \overline{U^{t}_{\,t}}(\rho) \equiv  \frac{1}{2\pi}
  \int_{-\infty}^{\infty} dt  \int_{-\pi}^{\pi} d\varphi\,\,
  U^{t}_{\,t}(t,\rho \cos \varphi,\rho \sin \varphi).
\end{equation}
It follows that
\begin{equation}\label{Gcon}
  \int_{{\cal R}^3} |det {\bf g}|^{\frac{1}{2}}
  G^{a}_{\,b}U^{b}_{\,a}{\bf \varepsilon}=
  \lim_{\epsilon \rightarrow 0} \pi \int_{\epsilon}^{\infty}
    d\rho \, \log H \frac{d}{d\rho}\left(
    \rho\frac{d}{d\rho}\overline{U^{t}_{\,t}} \right),
\end{equation}
where we have integrated by parts and imposed the condition
\begin{equation}
\lim_{\rho\rightarrow 0}\rho \frac{d}{d\rho}\log H=0.
\label{boundary}
\end{equation}

Now, let us assume that (\ref{disteins}) holds even for a less
well-behaved $H$. Let $\log H$ be the $C^1$-function given by
\begin{equation}
\log H= -8m\log \rho_{>},\label{H}
\end{equation}
where $\rho_{>}= max\{\rho, \xi \}$ with $\xi$ a constant
parameter which we will treat as a regulator. Eq.(\ref{H}) is a
solution of
\begin{equation}
\frac{d}{d\rho}(\rho\frac{d}{d\rho}\log H)=
-8m\delta(\rho-\xi),\label{greenH}
\end{equation}
that satisfies (\ref{boundary}) and where both sides of
(\ref{greenH}) should be understood as distributions on test
functions $\phi(\rho)\in \Phi({\cal R}^+)$. From (\ref{H}) it
follows
\begin{equation}
H(\rho)= (\rho_{>})^{-8m},\label{fH}
\end{equation}
which is also a well defined distribution on $\Phi({\cal R}^+)$.

From (\ref{Gcon}) and (\ref{H}) it follows
\begin{equation}\label{Gxi}
  \int_{{\cal R}^3} |det {\bf g}|^{\frac{1}{2}}
  G^{a}_{\,b}U^{b}_{\,a}{\bf \varepsilon}=
  -8\pi m \overline{U^{t}_{\,t}}(\xi),
\end{equation}
where we have integrated by parts and used (\ref{boundary}).

From (\ref{cinco}), (\ref{GR}) and (\ref{Gxi}), we have that the
Riemann tensor density is supported on
the surface $\rho = \xi$. Hence, the spacetime will be flat except
on this surface. Geometries of this kind in $(2+1)$-dimensional
gravity have been previously studied\cite{deser89,grignani}.

From (\ref{Gxi}), it follows
\begin{equation}
    \lim_{\xi \rightarrow 0}
  \int_{{\cal R}^3} |det {\bf g}|^{\frac{1}{2}}
  G^{a}_{\,b}U^{b}_{\,a}{\bf \varepsilon}=
  -8\pi m \overline{U^{t}_{\,t}}(0) =
  -8\pi m \int_{-\infty}^{\infty} dt\, U^{t}_{\,t}(t,0,0).
\end{equation}
Therefore, the $\xi \rightarrow 0$ limit of the density $|det {\bf
g}|^{\frac{1}{2}}{\bf G}$ is the distribution
\begin{equation}
\lim_{\xi \rightarrow 0}|det {\bf g}|^{\frac{1}{2}}G^a_{\,b}=
-8\pi m \delta^{(2)}_{(0)}\partial_t^{a}\, dt_b,\label{G-DJT}
\end{equation}
where $\delta^{(2)}_{(0)}$ is the usual two-dimensional Euclidean
$\delta$ distribution with support on the origin. This means that
the  spacetime $({\cal R}^3,{\bf g})$, where the metric ${\bf g}$
is given by (\ref{uno},\ref{fH}), can be identified in the $\xi
\rightarrow 0$ limit with the spacetime generated by a
distributional energy-momentum tensor ${\bf T}= T^t_{\,t} {\bf
d}t\otimes\partial_t$ with
\begin{equation}
|det {\bf g}|^{\frac{1}{2}}T^t_{\,t}= -m \delta^{(2)}_{(0)}.
\end{equation}
Since for $\rho >\xi$, the metrics (\ref{cero}) and
(\ref{uno},\ref{fH}) agree, a point source of mass $m$ at the
origin may be considered as the source for the metric
(\ref{cero}), as proposed in reference \cite{deser84}.

We now wish to make contact with the approach of reference
\cite{geroch}. The regularized metric (\ref{uno},\ref{fH})
is a continuous metric with a jump discontinuity of the
extrinsic curvature across the hypersurface of codimension
one $\rho= \xi$. Then the metric (\ref{uno},\ref{fH}) is a
regular metric $\forall \xi>0$. We have
\begin{equation}
{\bf g}= {\bf \eta} - (1-H)({\bf d}\rho\otimes{\bf d}\rho +
\rho^2{\bf d}\varphi\otimes{\bf d}\varphi)
\end{equation}
and
\begin{equation}
{\bf g}^{-1}= {\bf \eta}^{-1} - (1-H^{-1}) (
\partial_{\rho}\otimes
\partial_ {\rho} +
\rho^{-2}\partial_{\varphi}\otimes
\partial_ {\varphi}),
\end{equation}
where ${\bf \eta}$ is given by (\ref{minkowski}), $H$ is given by
(\ref{fH}) and we take for the differentiable structure that in
which $t, x=\rho \cos \varphi$ and $y=\rho \sin \varphi$ form a
smooth chart. Hence it follows that ${\bf g}$ and ${\bf g}^{-1}$
exist everywhere.

Let ${\bf U}$ be a test \ttensor{2}{0} field on ${\cal R}^3$. We
have
\begin{equation}
{\bf g}[{\bf U}]= \int_{{\cal R}^3}({\bf \eta}|{\bf
U})\omega_{\eta} - \int dt \int_0^{\xi}d\rho\int d\varphi\,\rho
(1-\xi^{-8m})(U^{xx} + U^{yy}) - \int dt
\int_{\xi}^{\infty}d\rho\int d\varphi\,\rho (1-\rho^{-8m})(U^{xx}
+ U^{yy}),
\end{equation}
where $\rho= \sqrt{x^2+y^2}$. Therefore, ${\bf g}$ is locally
bounded for $\xi>0$. Next, let ${\bf S}$ be a test \ttensor{0}{2}
field on ${\cal R}^3$. It follows that
\begin{equation}
{\bf g}^{-1}[{\bf S}]= \int_{{\cal R}^3}({\bf S}|{\bf
\eta}^{-1})\omega_{\eta} - \int dt \int_0^{\xi}d\rho\int
d\varphi\,\rho(1-\xi^{8m})(S_{xx} + S_{yy})-\int dt
\int_{\xi}^{\infty}d\rho\int d\varphi\,\rho(1-\rho^{8m})(S_{xx} +
S_{yy}).
\end{equation}
Hence, ${\bf g}^{-1}$ is locally bounded for all $\xi > 0$.
Finally, let ${\bf U}$ be a test \ttensor{3}{0} field on ${\cal
R}^3$. We find that the weak derivative in ${\bf \eta}$ of ${\bf
g}$ is given by
\begin{equation}
{\bf \nabla} {\bf g}[{\bf U}]= -{\bf g}[\eta \cdot {\bf \nabla}
{\bf U}]= \int_{{\cal R}^3} W_{cab}U^{cab}\omega_{\eta},
\end{equation}
where
\begin{equation}
W_{cab}= \cases{0, & $\rho<\xi$\cr
    -8m \rho^{-(1+8m)}d\rho_c(d\rho_ad\rho_b +
    \rho ^2d\varphi_a d\varphi_b), & $\rho>\xi$}.\label{weak2}
\end{equation}
It then follows that ${\bf \nabla}{\bf g}$ is locally square
integrable for $\xi > 0$. Therefore, (\ref{uno},\ref{fH}) is a
regular metric. Furthermore, since it is a continuous metric, we
have a well defined intrinsic $2$-geometry at the two-surface
$\rho=\xi$. It should be recalled that continuous regular metrics
can be suitably approximated by smooth metrics\cite{geroch}. In
this sense, the class of continuous regular metrics provides a
physically sensible idealization of the smooth metrics of general
relativity.

Now, from (\ref{uno},\ref{fH}) and (\ref{chris}) it follows that
\begin{equation}
C^c_{ab}= \cases{0, & $\rho<\xi$\cr
          \frac{4m}{\rho}\left(
          \partial_{\rho}^c d\rho_ad\rho_b + \partial_{\varphi}^c(d\rho_a
          d\varphi_b + d\rho_b d\varphi_a) -\rho^2
          \partial_{\rho}^c  d\varphi_a  d\varphi_b\right), & $\rho>\xi$}
          \label{chrisconreg}
\end{equation}
and
\begin{equation}
C^d_{m[b}C^m_{a]c}=0.\label{CCreg}
\end{equation}
From (\ref{RR}), (\ref{chrisconreg}) and (\ref{CCreg}) we have
\begin{equation}
R_{ab}= \nabla_{[c}C^c_{a]b}.
\end{equation}
Computations analogous to the previous ones give
\begin{equation}
R_{ab}[S^{ab}]= 4m \int dt\int d\varphi\,\left(S^{xx}(t,\xi\cos
\varphi ,\xi \sin\varphi) + S^{yy}(t,\xi\cos \varphi ,\xi
\sin\varphi)\right)\label{Riccireg}
\end{equation}
and
\begin{eqnarray}
G_{ab}[S^{ab}]&=& \nonumber R_{ab}[S^{ab}] \\ \nonumber&& -4m \int
dt\int d\varphi\,\left(-\xi^{8m}S^{tt}(t,\xi\cos \varphi ,\xi
\sin\varphi)+ S^{xx}(t,\xi\cos \varphi ,\xi \sin\varphi) +
S^{yy}(t,\xi\cos \varphi ,\xi \sin\varphi)\right)\\&=& 4m
\xi^{8m}\int dt\int d\varphi\,S^{tt}(t,\xi\cos \varphi ,\xi
\sin\varphi),
\end{eqnarray}
where in the last step we have used (\ref{Riccireg}). Hence
\begin{equation}
\lim_{\xi\rightarrow 0}R_{ab}[S^{ab}]=  8\pi m
\delta^{(2)}_{(0)}(dx_a dx_b +dy_a dy_b)[S^{ab}]
\end{equation}
and
\begin{equation}
\lim_{\xi\rightarrow 0}G_{ab}[S^{ab}]= 0,
\end{equation}
in agreement with (\ref{Riccifinal}) and (\ref{Gsemreg}).

For the sake of comparison, let us calculate $G^{a}_{\, b}$ for
the metric (\ref{uno},\ref{fH}). For a smooth metric we have
\begin{equation}
G^{a}_{\, b}= R^{a}_{\, b} -
\frac{1}{2}(g^{-1})^{cd}{\tilde{R}}_{cd}\delta^{a}_{\, b} +
(g^{-1})^{cd}C^e_{m[c}C^m_{e]d}\delta^{a}_{\, b} +
\nabla_{[c}\left(C^e_{e]d}(g^{-1})^{cd}\right)\delta^{a}_{\, b} +
C^e_{d[c}\nabla_{e]}(g^{-1})^{cd}\delta^{a}_{\, b},\label{ERM}
\end{equation}
where
\begin{equation}
R^{a}_{\, b}= (g^{-1})^{ac}{\tilde{R}}_{cb} + 2\nabla_{[c}\left(
C^{c}_{d]b}(g^{-1})^{ad}\right) +
2C^{c}_{b[c}\nabla_{d]}(g^{-1})^{ad} +
2(g^{-1})^{ad}C^{c}_{m[c}C^{m}_{d]b}.\label{RicciM}
\end{equation}
We take (\ref{ERM}) as the definition of the mixed Einstein tensor
for a regular metric. As a qualifying remark, it should be
recalled that for a regular metric, the outer product of any
number of metrics and inverse metrics with a single curvature
tensor can be interpreted as a distribution\cite{geroch}. Since
any contraction of a distribution is a distribution, for a regular
metric the Einstein tensors (\ref{ER}) and (\ref{ERM}) are well
defined as distributions.

Let ${\bf S}$ be a test \ttensor{1}{1} field on ${\cal R}^3$. From
(\ref{ERM}) and (\ref{RicciM}), a calculation analogous to that of
equation (\ref{Riccicalc}) leads to
\begin{equation}
R^{a}_{\, b}[S^{b}_{\, a}]= 4m\, \xi^{8m}\int dt\int
d\varphi\,\left(S^{x}_{\, x}(t,\xi\cos \varphi ,\xi \sin\varphi) +
S^{y}_{\, y}(t,\xi\cos \varphi ,\xi
\sin\varphi)\right)\label{RicciMreg}
\end{equation}
and
\begin{eqnarray}
G^{a}_{\, b}[S^{b}_{\, a}]&=& R^{a}_{\, b}[S^{b}_{\, a}] - 4m\,
\xi^{8m}\int dt\int d\varphi\,\left( S^{t}_{\, t}(t,\xi\cos
\varphi ,\xi \sin\varphi) + S^{x}_{\, x}(t,\xi\cos \varphi ,\xi
\sin\varphi) + S^{y}_{\, y}(t,\xi\cos \varphi ,\xi
\sin\varphi)\right)\nonumber \\&=&  - 4m\, \xi^{8m}\int dt\int
d\varphi\, S^{t}_{\, t}(t,\xi\cos \varphi ,\xi \sin\varphi),
\label{GMreg}
\end{eqnarray}
where we have used (\ref{RicciMreg}). Hence
\begin{equation}
\lim_{\xi\rightarrow 0}R^{a}_{\, b}[S^{b}_{\, a}]= 0
\end{equation}
and
\begin{equation}
\lim_{\xi\rightarrow 0}G^{a}_{\, b}[S^{b}_{\, a}]= 0.\label{GM0}
\end{equation}
A comparison of (\ref{G-DJT}) and (\ref{GM0}) illustrates the
dependence of (\ref{GM0}) on the volume element chosen. This is the
usual situation when tensors distributions are associated with
locally integrable tensors via Eq.(\ref{Tdist}), requiring the
specification of a reference volume element\cite{louko}. However, as
this example suggests, the approach that led us to (\ref{G-DJT})
appears to be more appropriate than the one used to arrive to
(\ref{Gsemreg}) and (\ref{GM0}), in order to relate the singularity
of the curvature of (\ref{cero}) to that of the Einstein tensor.

Finally, since general relativity is a covariant theory,
coordinate invariance of these results must be considered. The
present approach needs the introduction of a coordinate system
but, since the righthand side of (\ref{G-DJT}) is a tensor density
of weight one defined on ${\cal R}^2$, coordinate invariance of
the result is expected at least under those coordinate
transformations that do not involve the coordinate $t$.

\section{The Schwarzschild Geometry }\label{sec4}

The Schwarzschild solution is locally the only asymptotically flat,
spherically symmetric solution to the Einstein empty space field
equations. Schwarzschild spacetime is taken usually as that part of a
$4$-dimensional manifold with a metric of the form
\begin{equation}
{\bf g} =
    -(1-\frac{2m}{r}){\bf d}t\,{\bf d}t
    +(1-\frac{2m}{r})^{-1}{\bf d}r\,{\bf d}r
    + r^2({\bf d}\theta\,{\bf d}\theta
    +\sin ^2\theta{\bf d}\varphi\, {\bf d}\varphi).
    \label{sch}
\end{equation}
where $r > 2m$, $-\infty < t <\infty$, $0 \leq \theta \leq \pi$
and $0 \leq \varphi \leq 2\pi$. Following standard practice, in
(\ref{sch}) we have omitted writing the outer product sign.

Assuming that the manifold and (\ref{sch}) can be extended to include
the region $r\leq 2m$, we can ask for the source of this geometry.
The metric (\ref{sch}) does not fall within the class of regular
metrics\cite{geroch}. In the following, we will show that it is not a
semi-regular metric.

Let ${\bf \eta}$ be the ordinary Minkowski metric on ${\cal R}^4$
given by
\begin{equation}
{\bf \eta}
    = -{\bf d}t\,{\bf d}t +{\bf d}r\,{\bf d}r +
    r ^2({\bf d}\theta \,{\bf d}\theta +
    \sin^2 \theta{\bf d}\varphi \,{\bf d}\varphi).
    \label{minkowski4}
\end{equation}
We take for the differentiable structure that in which $t, x=r
\cos \varphi\sin \theta, y=r \sin \varphi\sin \theta$ and $z=r\cos
\theta$ form a smooth chart. It follows that
\begin{equation}
{\bf g}= {\bf \eta} + \frac{2m}{r}{\bf d}t\,{\bf d}t -
\frac{2m}{2m-r}{\bf d}r\,{\bf d}r
\end{equation}
and
\begin{equation}
{\bf g}^{-1}= {\bf \eta}^{-1} +
\frac{2m}{2m-r}\partial_t\,\partial_t -
\frac{2m}{r}\partial_r\,\partial_r,
\end{equation}
where $r=\sqrt{x^2 + y^2 + z^2}$. We have that ${\bf g}$ and ${\bf
g}^{-1}$ exist almost everywhere. Let ${\bf U}$ be a test
\ttensor{2}{0} field on ${\cal R}^4$ with support on $r<2m$. Thus,
\begin{equation}
{\bf g}[{\bf U}]= \int_{{\cal R}^4}({\bf \eta}|{\bf
U})\omega_{\eta} - \int_{{\cal R}^3}(\frac{2m}{r}\,U^{tt} -
\frac{2m}{2m-r}dr_{a}dr_{b}U^{ab})\omega_{\eta}.
\end{equation}
Therefore, ${\bf g}$ is locally integrable. Now, let ${\bf S}$ be
a test \ttensor{0}{2} field on ${\cal R}^4$ with support on
$r<2m$. It follows that
\begin{equation}
{\bf g}^{-1}[{\bf S}]= \int_{{\cal R}^4}({\bf S}|{\bf
\eta}^{-1})\omega_{\eta} - \int_{{\cal
R}^4}(\frac{2m}{2m-r}\,S_{tt} -
\frac{2m}{r}\partial_r^{a}\partial_r^{b}\,S_{ab})\omega_{\eta}.
\end{equation}
Hence, ${\bf g}^{-1}$ is locally integrable. Next, let ${\bf U}$
be a test \ttensor{3}{0} field on ${\cal R}^4$ with support on
$r<2m$. The weak derivative in ${\bf \eta}$ of ${\bf g}$ exist
almost everywhere and is given by
\begin{equation}
{\bf \nabla} {\bf g}[{\bf U}]= -{\bf g}[\eta \cdot {\bf \nabla} {\bf
U}]= \lim_{\epsilon \rightarrow 0} \int_{r >\epsilon} g_{ab}\nabla_c
U^{cab} \omega_{\eta}= \int_{{\cal R}^4} W_{cab}U^{cab}\omega_{\eta}
\end{equation}
where
\begin{eqnarray}
W_{cab}= &-&\frac{2m}{r^2}dr_{c}dt_{a}dt_{b} -
\frac{2m}{(24-r)^2}dr_{c}dr_{a}dr_{b} \nonumber\\&-&
\frac{1}{r}\frac{2m}{2m-r}\left(r^2(d\theta_{c}d\theta_{a} +
\sin^2\theta d\varphi_{c}d\varphi_{a})dr_{b} +
r^2dr_{a}(d\theta_{c}d\theta_{b} + \sin^2\theta
d\varphi_{c}d\varphi_{b})\right) .\label{weaksch}
\end{eqnarray}
It then follows that ${\bf \nabla}{\bf g}$ is locally integrable.
However, it is not locally square-integrable in $r<2m$ due to the
fact that $-\frac{2m}{r^2}dr_{c}dt_{a}dt_{b}$ is not locally
square-integrable. Finally, from (\ref{chris}) and (\ref{weaksch}) we
have
\begin{eqnarray}
C^{c}_{ab}= &-&\frac{1}{r}\frac{m}{2m-r}\partial^{c}_t(dr_adt_b +
dt_adr_b) + \frac{1}{r}\frac{m}{2m-r}\partial^{c}_rdr_adr_b
\nonumber\\&+& \frac{2m}{r^2}\partial^{c}_r(r^2d\theta_a d\theta_b
+ r^2\sin^2\theta d\varphi_a d\varphi_b) - (\frac{2m}{r^3} -
\frac{m}{r^2})\partial^{c}_r dt_a dt_b,
\end{eqnarray}
which is not locally integrable because
$\frac{2m}{r^3}\partial^{c}_r dt_a dt_b$ is not locally
integrable. Therefore the metric (\ref{sch}) is not a semi-regular
metric.

In these case, however, distributional curvature and Einstein
tensors can be obtained for (\ref{sch}) through regularization
procedures. Following Ref.\cite{balasin93}, consider a metric of
the form
\begin{equation}
{\bf g}
    = h{\bf d}t\,{\bf d}t - h^{-1}{\bf d}r\,{\bf d}r
    +r^2({\bf d}\theta \,{\bf d}\theta +
    \sin^2\theta {\bf d}\varphi \,{\bf d}\varphi),
    \label{nueve}
\end{equation}
where $0 \leq r < 2m$ and $h$ is a $C^{\infty}$ function of $r$.
The evaluation of the Einstein tensor proceeds now in a
straightforward manner. We omit the details of the calculation;
the result is
\begin{equation}
G  = -\left[\frac{1}{r^2}\frac{d}{dr}(r(1+h))\right]
   ({\bf d}t{\partial}_t + {\bf d}r{\partial}_r)
    -\frac{1}{2}\left[\frac{1}{r^2}\frac{d}{dr}(r^2\frac{d}{dr}h)\right]
   ({\bf d}\theta{\partial}_\theta
    + {\bf d}\varphi {\partial}_\varphi).
    \label{once}
\end{equation}
To find a distributional source for the Schwarzschild spacetime,
consider the components of (\ref{once}) as distributions on the
space of test functions $\phi \in {\cal D}({\cal R}^3)$ with
support on the region $r< 2m$, where ${\cal R}^3$ is endowed with
the usual $C^{\infty}$ Euclidean metric, and $h$ given by
\begin{equation}
h(r)=-1 +\frac{2m}r f, \label{curvreg}
\end{equation}
where $f= f_{\lambda}(r)$ is a $C^{\infty}$ function such that
$f_{\lambda}(0)= 0$ and satisfying $\lim_{\lambda \rightarrow
\lambda_0} f_{\lambda}(r)\rightarrow 1$. The metric
(\ref{nueve},\ref{curvreg}) is a regularized version of
(\ref{sch}). From Einstein field equations and
(\ref{once},\ref{curvreg}), a source ${\bf T}_f$ which depends
explicitly on the regularization function $f_{\lambda}$ is
obtained. By taking $f_{\lambda}(r)= r^{\lambda}$, the
$\lambda\rightarrow 0$ distributional limit, gives
\begin{equation}
{\bf T}
    = -m\delta ^{(3)}_{(0)}[ {\bf d}t{\partial}_t
    +{\bf d}r{\partial}_r-\frac{1}{2}{\bf d}\theta{\partial}_\theta
    -\frac{1}{2}{\bf d}\varphi {\partial}_\varphi ],  \label{ge}
\end{equation}
where $\delta^{(3)}_{(0)}$ is the usual three-dimensional Euclidean
$\delta$ distribution. By virtue of the distributional identity in
${\cal R}^3$
\begin{equation}
\delta ^{(3)}_{(0)}[{\bf d}r{\partial}_r
    -\frac{1}{2}{\bf d}\theta{\partial}_\theta
    -\frac{1}{2}{\bf d}\varphi {\partial}_\varphi ]=0,\label{identity}
\end{equation}
it follows that (\ref{ge}) has the simpler expression
\begin{equation}
{\bf T}= -m\delta ^{(3)}_{(0)}{\bf d}t{\partial}_t.
\label{endresult}
\end{equation}
The distributional tensor $\bf T$ given by (\ref{endresult}),
vanishes for $m=0$ and has support on the region $|\vec{x}|= 0$.
Therefore, the Schwarzschild geometry can be considered as generated
by the distributional energy-momentum tensor
(\ref{endresult})\cite{balasin93}. It should be noted that
(\ref{endresult}) can also be obtained without the need of
regularization procedures, approximating (\ref{sch}) by analytic
metrics and using the fact that reciprocals of analytic functions
provide well-defined distributions\cite{parker}.

We shall now show that the regularization (\ref{curvreg}) with
$f_{\lambda}(r)= r^{\lambda}$ does not provide a regular metric in
the sense of reference \cite{geroch}. Let ${\bf U}$ and ${\bf S}$ be
a test \ttensor{2}{0} and \ttensor{0}{2} fields on ${\cal R}^4$,
respectively, with support on $r<2m$. We have
\begin{equation}
({\bf g}|{\bf U})= ({\bf \eta}|{\bf U}) -
(\frac{2mr^{\lambda}}{r}\,U^{tt} -
\frac{2mr^{\lambda}}{2mr^{\lambda}-r}dr_{a}dr_{b}U^{ab})
\end{equation}
and
\begin{equation}
({\bf g}^{-1}|{\bf S})= ({\bf S}|{\bf \eta}^{-1}) -
(\frac{2mr^{\lambda}}{2mr^{\lambda}-r}\,S_{tt} -
\frac{2mr^{\lambda}}{r}\partial_r^{a}\partial_r^{b}\,S_{ab}).
\end{equation}
It then follows that ${\bf g}$ and ${\bf g}^{-1}$, although
locally integrable for $\lambda \geq 0$, are locally bounded for
$\lambda \geq 1$. Further, the weak derivative in ${\bf \eta}$ of
${\bf g}$ exist almost everywhere and is given by
\begin{equation}
{\bf \nabla} {\bf g}[{\bf U}]= -{\bf g}[\eta \cdot {\bf \nabla}
{\bf U}]= \int_{{\cal R}^4} W_{cab}U^{cab}\omega_{\eta},
\end{equation}
where
\begin{equation}
W_{cab}= -(1-\lambda)2m r^{\lambda - 2}dr_{c}dt_{a}dt_{b} -
2m\nabla_c\left(\frac{r^{\lambda}}{2mr^{\lambda}-r}dr_{a}dr_{b}\right).
\end{equation}
The weak derivative is locally integrable for $\lambda \geq 0$ and
locally square integrable for $\lambda \geq \frac{1}{2}$. It then
follows that this regularization does not provide a regular metric
$\forall \lambda > 0$. Thus the distributional meaning of the
Riemann tensor and its contractions, in the sense of reference
\cite{geroch}, is questionable at the intermediate steps of the
calculation.

We note in passing that (\ref{endresult}) may be obtained with
different regularization functions
$f_{\lambda}$\cite{balasin93,kawai}. However, the invariance of
the regularization procedure is uncertain and (\ref{endresult}) is
obtained only by imposing some rather {\it ad hoc} regularization
prescriptions\cite{kawai}. Using the Kerr-Schild {\it ansatz} for
the metric tensor and assuming that under the regularization
procedure the metric maintains its Kerr-Schild form, it has been
shown that (\ref{endresult}) can also be obtained\cite{balasin94}.

We turn now to the problem posed by Einstein field equations ${\bf
G}= 8\pi {\bf T}$, with ${\bf T}$ given by (\ref{endresult}). What
we want to address here is how to arrive to (\ref{endresult}) by a
regularization procedure guaranteeing that the curvature tensor
and its contractions be well defined tensor distributions at all
intermediate steps of the calculation. The regularization
procedure discussed in the previous section can be applied here.
As we have shown, within this approach many of the difficulties
arising from lack of invariance or caused by regularization
ambiguities are avoided, or at least, deferred.

We take (\ref{once}) as a definition of the Einstein tensor. Let
${\bf U}$ be a test \ttensor{1}{1} field on ${\cal R}^4$. Next,
consider the mixed-index Einstein tensor density $|det {\bf
g}|^{\frac{1}{2}}{\bf G}$, with $|det {\bf g}|^{\frac{1}{2}}$ the
density generated by the metric {\bf g} in (\ref{nueve}), as a
functional on the space of test tensor fields through
\begin{equation}
  \int_{{\cal R}^4} |det {\bf
  g}|^{\frac{1}{2}}G^{a}_{\,b}U^{b}_{\,a}{\bf \varepsilon}
  = \int_{{\cal R}^4} dt dx dy dz\  G^{a}_{\,b}U^{b}_{\,a},
\end{equation}
where it is understood that $G^{a}_{\,b}$ and $U^{b}_{\,a}$ are
Cartesian components as functions of Cartesian coordinates
$(t,x,y,z)$ with $x=r\cos\varphi\sin\theta$, $y=r\sin\varphi\sin\theta$
and $z=r\cos\theta$. Then
\begin{eqnarray}
  \int_{{\cal R}^4} |det {\bf
  g}|^{\frac{1}{2}}G^{a}_{\,b}U^{b}_{\,a}{\bf \varepsilon}
  = \lim_{\epsilon \rightarrow  0} 4\pi \int_{r>\epsilon} & dr & \left[
  - \frac{d}{dr} (r(1+h))\right] \left( \overline{U^{t}_{\,t}}(r)
  + \overline{\partial^{a}_{\,r} dr_b U^{b}_{\,a}}(r)
  \right)\nonumber \\
  &&
  \ +\frac{1}{2} \left[- \frac{d}{dr} (r^2\frac{d}{dr} h) \right]
  \overline{(\partial^{a}_{\,\theta} d\theta_b +
  \partial^{a}_{\,\varphi}d\varphi_b)U^{b}_{\,a}}(r),
\end{eqnarray}
where
\begin{equation}
    \overline{U^{t}_{\,t}}(r) \equiv \frac{1}{4\pi}
    \int_{-\infty}^{\infty} dt \int_{0}^{\pi} d\theta
    \int_{0}^{2\pi} d\varphi\, \sin\theta \, U^{t}_{\,t},
\end{equation}
\begin{eqnarray}
  \overline{\partial^{a}_{\,r} dr_b U^{b}_{\,a}}(r) \equiv
  \frac{1}{4\pi} \int_{-\infty}^{\infty} dt \int_{0}^{\pi} d\theta
    \int_{0}^{2\pi} d\varphi\, \sin\theta && \left[ \sin^2\theta
    (\cos^2\varphi\, U^{x}_{\,x}+ \sin^2\varphi\, U^{y}_{\,y}) +
    \cos^2\theta\, U^{z}_{\,z} \right. \nonumber \\
    && + \sin^2\theta \sin\varphi\cos\varphi (U^{y}_{\,x} + U^{x}_{\,y})\nonumber\\
    && + \sin\theta \cos\theta \sin\varphi(U^{y}_{\,z} + U^{z}_{\,y}) \nonumber \\
    && \left. + \sin\theta \cos\theta \cos\varphi
    (U^{z}_{\,x} + U^{x}_{\,z}) \right]
\end{eqnarray}
and
\begin{eqnarray}
  \overline{(\partial^{a}_{\,\theta} d\theta_b +
  \partial^{a}_{\,\varphi}d\varphi_b)U^{b}_{\,a}}(r) \equiv
  \frac{1}{4\pi} \int_{-\infty}^{\infty} dt \int_{0}^{\pi} d\theta
    \int_{0}^{2\pi} d\varphi\, \sin\theta && \left[ \cos^2\theta
    (\cos^2\varphi\, U^{x}_{\,x}+ \sin^2\varphi\, U^{y}_{\,y}) +
    \sin^2\theta\, U^{z}_{\,z} + \right. \nonumber\\
    && + \sin^2\varphi\, U^{x}_{\,x} + \cos^2\varphi\, U^{y}_{\,y}\nonumber \\
    && - \sin^2\theta \sin\varphi\cos\varphi (U^{y}_{\,x} + U^{x}_{\,y}) \nonumber \\
    && - \sin\theta \cos\theta \sin\varphi (U^{y}_{\,z} + U^{z}_{\,y}) \nonumber \\
    && \left. - \sin\theta \cos\theta \cos\varphi
    (U^{z}_{\,x} + U^{x}_{\,z}) \right].
\end{eqnarray}

It follows
\begin{equation}\label{Gsch}
  \int_{{\cal R}^4} |det {\bf
  g}|^{\frac{1}{2}}G^{a}_{\,b}U^{b}_{\,a}{\bf \varepsilon}
  = \lim_{\epsilon \rightarrow 0} \left\{ 4\pi
  \int_{\epsilon}^{\infty} dr\, r(1+h) \frac{d}{dr} \left[
  \overline{U^{t}_{\,t}} + \overline{\partial^{a}_{\,r} dr_b\,
  U^{b}_{\,a}}\right] - 2\pi \int_{\epsilon}^{\infty} h
  \frac{d}{dr} \left[r^2 \frac{d}{dr}
  \overline{(\partial^{a}_{\theta} \theta_b + \partial^{a}_{\varphi}
  d\varphi_b) U^{b}_{\,a}}\right]\right\},
\end{equation}
where we have integrated by parts an imposed the condition
\begin{equation}
\lim_{r\rightarrow 0}r^2 \frac{d}{dr}h=0. \label{boundarysch}
\end{equation}

Now, let us assume that (\ref{Gsch}) holds even for a less
well-behaved $h$. Let $h$ be the $C^1$-function given by
\begin{equation}
h= -1 + \frac{2m}{r_{>}},\label{h}
\end{equation}
where $r_{>}= max\{r, \xi \}$ with $0 < \xi < 2m$ a constant
parameter which we will treat as a regulator. Eq.(\ref{h}) is the
solution of
\begin{equation}
\frac{d}{dr}(r^2\frac{d}{dr}h)= -2m\delta(\rho-\xi),\label{greenh}
\end{equation}
where both sides should be understood as distributions on test
functions $\phi(r)\in \Phi({\cal R}^+)$ whose support is
contained in $r < 2m$. Further, (\ref{h}) satisfies
(\ref{boundarysch}) and
\begin{equation}
h\mid_{r=2m}=0,\label{border}
\end{equation}
where the condition (\ref{border}) ensures continuity of the
metric tensor at $r = 2m$, assuming that (\ref{sch}) holds for $r
> 2m$. Note that, although for $r < 2m$ the coordinate $r$ is
timelike and the spacetime is no static, causality with respect to
$r$ cannot be used as a criterion to obtain $h$.

From (\ref{Gsch}) and (\ref{h}) it follows

\begin{equation}\label{Gttxisch}
  \int_{{\cal R}^4} |det {\bf
  g}|^{\frac{1}{2}}G^{a}_{\,b}U^{b}_{\,a}{\bf \varepsilon}
  = -8\pi m \frac{1}{\xi} \int_{0}^{\xi} dr\, \left[
  \overline{U^{t}_{\,t}} + \overline{\partial^{a}_{\,r} dr_b\,
  U^{b}_{\,a}}\right] + 4\pi m
  \overline{(\partial^{a}_{\theta} \theta_b + \partial^{a}_{\varphi}
  d\varphi_b) U^{b}_{\,a}}(\xi).
\end{equation}
Eq. (\ref{Gttxisch}) implies that the density $|det {\bf
g}|^{\frac{1}{2}}{\bf G}$  has support on the four dimensional
submanifold $r<\xi$.

Now, from (\ref{Gttxisch}) we have
\begin{equation}
  \lim_{\xi \rightarrow 0} \int_{{\cal R}^4} |det {\bf
  g}|^{\frac{1}{2}}G^{a}_{\,b}U^{b}_{\,a}{\bf \varepsilon}
  = -8\pi m \left[
  \overline{U^{t}_{\,t}}(0) + \overline{\partial^{a}_{\,r} dr_b\,
  U^{b}_{\,a}}(0)\right] + 4\pi m
  \overline{(\partial^{a}_{\theta} \theta_b + \partial^{a}_{\varphi}
  d\varphi_b) U^{b}_{\,a}}(0),
\end{equation}
where
\begin{equation}
  \overline{\partial^{a}_{\,r} dr_b\, U^{b}_{\,a}}(0) =
  \frac{1}{3} \int_{-\infty}^{\infty} dt \,
  \left(U^{x}_{\,x}(t,0,0,0) + U^{y}_{\,y}(t,0,0,0) + U^{z}_{\,z}(t,0,0,0) \right)
\end{equation}
and

\begin{equation}
  \overline{(\partial^{a}_{\theta} \theta_b + \partial^{a}_{\varphi}
  d\varphi_b) U^{b}_{\,a}}(0) =
  \frac{2}{3} \int_{-\infty}^{\infty} dt \,
  \left(U^{x}_{\,x}(t,0,0,0) + U^{y}_{\,y}(t,0,0,0) + U^{z}_{\,z}(t,0,0,0)
  \right).
\end{equation}
It follows
\begin{equation}
  \lim_{\xi \rightarrow 0} \int_{{\cal R}^4} |det {\bf
  g}|^{\frac{1}{2}}G^{a}_{\,b}U^{b}_{\,a}{\bf \varepsilon}=
  -8\pi m \int_{-\infty}^{\infty} dt\,U^{t}_{\,t}(t,0,0,0).
\end{equation}
Therefore, the $\xi\rightarrow 0$ limit of the density $|det {\bf
g}|^{\frac{1}{2}}{\bf G}$ is the distribution
\begin{equation}\label{same1}
   \lim_{\xi\rightarrow 0} |det {\bf g}|^{\frac{1}{2}}{\bf G} =
   -8\pi m \delta^{(3)}_{(0)} {\bf d}t\, \partial t
\end{equation}
where $\delta^{(3)}_{(0)}$ is the usual three-dimensional
Euclidean $\delta$ distribution with support on the origin. This
means that the spacetime $({\cal R}^4,{\bf g})$, where the metric
${\bf g}$ is given by (\ref{nueve},\ref{h}), can be identified in
the $\xi \rightarrow 0$ limit with the spacetime generated by the
distributional energy-momentum tensor ${\bf T}$ for which
\begin{equation}
|det {\bf g}|^{\frac{1}{2}}{\bf T}
     = -m\delta ^{(3)}_{(0)}{\bf d}t{\partial}_t.
\end{equation}

Since for $r >\xi$, the metrics (\ref{sch}) and (\ref{nueve}) with
$h$ given by (\ref{h}) coincide, a point source of mass $m$ at the
origin may be considered as the source for the Schwarzschild
geometry, as claimed in references \cite{parker,balasin93}. Note that
since the righthand side of (\ref{same1}) is a tensor density of
weight one defined on ${\cal R}^3$, coordinate invariance of this
result is expected at least under those coordinate transformations
that do not involve the coordinate $t$.

As follows from (\ref{h}), $h$ is continuous at $r=\xi$ .
Therefore, the usually required continuity of the metric is
ensured and the intrinsic 3-geometry of the hypersurface $r=\xi$ is
well defined. On the other hand, the extrinsic curvature is
discontinuous across this surface of codimension one. It then
follows that the metric (\ref{nueve},\ref{h}) is a
regular metric $\forall\xi>0$. We have
\begin{equation}
{\bf g}= {\bf \eta} + \frac{2m}{r_{>}}{\bf d}t\,{\bf d}t -
\frac{2m}{2m-r_{>}}{\bf d}r\,{\bf d}r
\end{equation}
and
\begin{equation}
{\bf g}^{-1}= {\bf \eta}^{-1} +
\frac{2m}{2m-r_{>}}\partial_t\,\partial_t -
\frac{2m}{r_{>}}\partial_r\,\partial_r.
\end{equation}
Thus ${\bf g}$ and ${\bf g}^{-1}$ exist almost everywhere and are
locally bounded $\forall \xi >0$. Let ${\bf U}$ be a test
\ttensor{3}{0} field on ${\cal R}^4$ with support on $r<2m$. The
weak derivative in ${\bf \eta}$ of ${\bf g}$ exist almost
everywhere and is given by
\begin{equation}
{\bf \nabla} {\bf g}[{\bf U}]= \int_{{\cal R}^4}
W_{cab}U^{cab}\omega_{\eta},
\end{equation}
where
\begin{equation}
W_{cab}= \cases{-
\frac{1}{r}\frac{2m}{2m-\xi}\left(r^2(d\theta_{c}d\theta_{a} +
\sin^2\theta d\varphi_{c}d\varphi_{a})dr_{b} +
r^2dr_{a}(d\theta_{c}d\theta_{b} + \sin^2\theta
d\varphi_{c}d\varphi_{b})\right),& $r<\xi$\cr \cr
-\frac{2m}{r^2}dr_{c}dt_{a}dt_{b} -
\frac{2m}{(24-r)^2}dr_{c}dr_{a}dr_{b}\cr \qquad -
\frac{1}{r}\frac{2m}{2m-r}\left(r^2(d\theta_{c}d\theta_{a} +
\sin^2\theta d\varphi_{c}d\varphi_{a})dr_{b} +
r^2dr_{a}(d\theta_{c}d\theta_{b} + \sin^2\theta
d\varphi_{c}d\varphi_{b})\right),& $r>\xi.$}
\end{equation}
This weak derivative is locally square-integrable. Hence,
(\ref{nueve},\ref{h}) is a regular metric as expected.

Computations analogous to the previous ones show that
\begin{eqnarray}\label{f1}
  R_{ab}[U^{ab}] &=&
  m \int_{r=\xi} \frac{1}{\xi^2} \left(
  (1-\frac{2m}{\xi}) dt_a  dt_b -
  (1-\frac{2m}{\xi})^{-1} dr_a  dr_b \right)U^{ab} \sigma
  \nonumber\\
  &&+
  \frac{2m}{\xi} \int_{-\infty}^\infty dt \int_0^\xi dr \int_0^\pi
  d\theta \int_0^{2\pi}d\varphi \, \sin\theta\, (r^2 d\theta_a
  d\theta_b + r^2\sin^2\theta d\varphi_a d\varphi_b) U^{ab},
\end{eqnarray}
\begin{eqnarray}\label{f2}
  G_{ab}[U^{ab}] &=&
  \frac{2m}{\xi} \int_{-\infty}^\infty dt \int_0^\xi dr \int_0^\pi
  d\theta \int_0^{2\pi}d\varphi \, \sin\theta \,\left(
  (1-\frac{2m}{\xi}) dt_a  dt_b  -
  (1-\frac{2m}{\xi})^{-1} dr_a  dr_b  \right)U^{ab} \nonumber \\
  &&- 2m \int_{r=\xi} \frac{1}{\xi^2} (r^2 d\theta_a
  d\theta_b + r^2\sin^2\theta d\varphi_a d\varphi_b) U^{ab}\sigma,
\end{eqnarray}
\begin{eqnarray}\label{f3}
  R^a_{\, b}[U^b_{\, a}] &=&
  - \int_{r=\xi} \frac{1}{\xi^2} (\partial^a_t dt_b
  + \partial^a_r dr_b) U^b_{\, a}\sigma \nonumber \\
  &&+
  \frac{2m}{\xi} \int_{-\infty}^\infty dt \int_0^\xi dr \int_0^\pi
  d\theta \int_0^{2\pi}d\varphi \, \sin\theta \,(\partial^a_\theta
  d\theta_b + \partial^a_\varphi d\varphi_b)U^b_{\, a}
\end{eqnarray}
and
\begin{eqnarray}\label{f4}
  G^a_{\, b}[U^b_{\, a}] &=&
  - \frac{2m}{\xi} \int_{-\infty}^\infty dt \int_0^\xi dr
  \int_0^\pi
  d\theta \int_0^{2\pi}d\varphi \, \sin\theta \,(\partial^a_t dt_b
  + \partial^a_r dr_b) U^b_{\, a} \nonumber \\
  &&+
  m \int_{r=\xi} \frac{1}{\xi^2} (\partial^a_\theta
  d\theta_b + \partial^a_\varphi d\varphi_b)U^b_{\, a}\sigma,
\end{eqnarray}
where it is understood that $U^{ab}$ and $U^{a}_{\,b}$ are
Cartesian components as functions of Cartesian coordinates.

From (\ref{f1}-\ref{f4}) it follows that we have well defined
distributional Ricci and Einstein tensors $\forall \xi >0$.
However, in the $\xi \rightarrow 0$ limit, only the mixed
components of these tensors turn out to be well defined
distributions,
\begin{equation}
\lim_{\xi \rightarrow 0}R^a_{\, b}[U^b_{\, a}]= 4\pi m
\delta^{(3)}_{(0)}\left[-{\partial}^{a}_t dt_{b}
    + {\partial}^{a}_x dx_b
    + {\partial}^{a}_y dy_{b}
    + {\partial}^{a}_z dz_{b} \right][U^b_{\, a}]
\end{equation}
and
\begin{equation}
\lim_{\xi \rightarrow 0}G^a_{\, b}[U^b_{\, a}]=
 -8\pi m\delta ^{(3)}_{(0)}{\partial}^{a}_tdt_{b}[U^b_{\,
    a}].\label{same2}
\end{equation}

Notice that the result (\ref{same1}) is recovered. In this case,
both approaches give the same result since the volume element of
the local auxiliary metric (\ref{minkowski4}) agrees with the
volume element of the regularized metric (\ref{nueve},\ref{h}).

\section{Concluding Remarks}
We have proposed a particular kind of regularization for the
metrics of two singular spacetimes: the $2+1$-dimensional
spacetime around a massive point source and the Schwarzschild
spacetime. In these two rather different examples, we have shown
that a satisfactory regularization procedure is obtained by
requiring that (i) the density $|det {\bf
g}|^{\frac{1}{2}}G^a_{\,b}$, associated to the Einstein tensor
$G^a_{\,b}$ of the regularized metric, rather than the Einstein
tensor itself, be a distribution and (ii) the regularized metric
be a continuous metric with a discontinuous extrinsic curvature
across a non-null hypersurface of codimension one. For these
examples, the regularized metrics are regular metrics
with well defined distributional curvature and Einstein tensors at
all the intermediate steps of the calculation. The distributional
limit, when the regularization is removed, assigns a well defined
distribution to the mixed-index Einstein tensor density with
support on the singularity. Further, some interesting
relationships between the curvatures obtained in different
approaches have been put forward. We have not discussed
exhaustively the coordinate invariance of the results. This
important issue will be considered elsewhere. But the strong
constraints that the present approach imposes on the
regularization procedure suggest that it may be useful to get
further insight about the distributional meaning of the curvature
of non-regular metrics.

\end{document}